\documentclass[a4paper,showpacs,preprintnumbers,amsmath,amssymb,nofootinbib,floatfix]{revtex4}
\def \be {\begin{equation}} 

\def \ee {\end{equation}} 
\def \bea {\begin{eqnarray}} 
\def \eea {\end{eqnarray}} 

\usepackage{graphicx}

\newcommand*{\ltsim}{\ {\raise-.75ex\hbox{$\buildrel<\over\sim$}}\ }
\newcommand*{\gtsim}{\ {\raise-.75ex\hbox{$\buildrel>\over\sim$}}\ }
\newcommand*{\proptosim}{\ {\raise-.75ex\hbox{$\buildrel\propto\over\sim$}}\ }

\begin{document}
\title{Reconstruction of the Cosmic Equation of State for High Redshift}

\author{A. M. Velasquez-Toribio}
\email{alan.toribio@ufes.br}
\affiliation{Departamento de F\'{\i}sica, Universidade Federal do Espirito Santo,  29075-910 Vit\'{\o}ria - ES, Brasil}

\author{M. M. Machado}
\email{marlonmoro92@googlemail.com }
\affiliation{Departamento de F\'{\i}sica, Universidade Federal do Espirito Santo,  29075-910 Vit\'{\o}ria - ES, Brasil}

\author{Julio C. Fabris}
\email{julio.fabris@cosmo-ufes.org}
\affiliation{Departamento de F\'{\i}sica, Universidade Federal do Espirito Santo,  29075-910 Vit\'{\o}ria - ES, Brasil}
\affiliation{National  Research  Nuclear  University  MEPhI,  Kashirskoe  sh.   31,  Moscow  115409, Russia}

\date{\today}

\begin{abstract}
We investigate the possibilities of reconstructing the cosmic equation of state (EoS) for high redshift.
In order to obtain general results, we use two model-independent approaches.
The first reconstructs the EoS using comoving distance and the second makes use of the Hubble parameter data.
To implement the first method, we use a recent set of Gamma-Ray Bursts (GRBs) measures.
To implement the second method, we generate simulated data using the Sandage-Loeb ($SL$) effect;
for the fiducial model, we use the $\Lambda CDM$ model.
In both cases, the statistical analysis is conducted through the Gaussian processes (non-parametric).
In general, we demonstrate that this methodology for reconstructing the EoS using a non-parametric method 
plus a model-independent approach works appropriately due to the feasibility of calculation and the 
ease of introducing a priori information ($H_ {0}$ and $\Omega_{m0}$).
In the near future, following this methodology with a higher number of high quality data 
will help obtain strong restrictions for the EoS. 

\end{abstract}
\maketitle

\section{Introduction}

The accelerated expansion of the universe is one of the biggest problems of cosmology 
today. Initially, it was associated with a cosmological constant or vacuum energy and, 
subsequently, models with scalar fields (also known as quintessence models) were evoked. Other possibilities include modified gravitation, 
extra dimensions, and so on. For a recent review, see references \cite{pebbles,mortonson,clifton}.

On the other hand, among the different cosmological observables, the cosmic equation of state (EoS) is of fundamental importance, 
as it carries the kinematic and dynamic information of a given cosmological model.
The reconstruction of these observables has been widely considered in the literature using different types of 
cosmological data, such as the following: Supernovae Ia, cosmic background radiation, clusters of galaxies, baryon acoustic oscillations ($BAO$), Hubble parameter data, 
$f\sigma_{8}$, and so on. Additionally, different statistical reconstruction techniques have been used  \cite{susuki, aghanim,anagnostopoulos}.

Nevertheless, the reconstruction of this observable has not been considered for high redshift, in principle, due to the lack of
data for any redshift greater than $2.0$. However, this question is currently changing and we can consider the reconstruction of the EoS ($w(z)$) for high redshifts.
Understanding in detail how $w(z)$ evolves as a function of time is fundamental to know the nature of dark energy. For instance, to answer the question
if the accelerated expansion is consistent with a local inhomogeneous effect or if the accelerated expansion is a phase of the evolution of the whole universe,
we must to know in detail the form of the EoS.

In this paper we reconstruct $w(z)$ using two model-independent approaches proposed in the literature \cite{zunckel,metodo,seikel,huterer}.
Specifically, these approximations are given by equations (6) and (17). 
The reconstruction of the EoS is not completely 
independent of the model because we need to assume explicit values
of $\Omega_{m0} $ and $H_{0}$. In general, there is no way, in cosmology, 
to determine the EoS in a completely model-independent way.
However, the choice of an adequate statistical method can help to enhance the model-independent approach. 
In that regard, we use the method of the Gaussian process as a statistical method, 
which is a non-parametric method and does not require knowing a specific parametric form of the EoS.
This statistical approach allows us to get closer to the ideal of a reconstruction that is completely independent from the model.

If the EoS can be determined with high precision, then we can discriminate between different types of cosmological models. However,
in the present paper, due to the quality and quantity of the current data, we do not get strong 
constraints to discriminate between models. Nevertheless, we prove that it is possible to use measured and simulated data to reconstruct 
$w(z)$ for high redshift. The main message of the article is that, even with the inconclusive estimates with the current data, our proposed methodology may be very useful for future data provided by some ongoing projects, such as the Euclid or the LSST among others\footnote{Information about the project Euclid, see the web page:https://www.euclid-ec.org/ and on the LSST:https://www.lsst.org/}.
  
Our paper is organized as follows. In Section II, we present the two reconstruction methods of the EoS and the Gaussian processes.
Section III is devoted to our results and in the Section IV, we present the conclusions of our study.

\section{Reconstruction of Equation of State}
We use two methods to reconstruct the EoS: the first method makes use of distance measurements from Gamma-Ray Bursts (GRBs) 
and the second uses simulated data of the Hubble parameter generated by the Sandage-Loeb (SL) effect. 
In this manner, we can form a reconstruction based on observed data and another on simulated data.
Both methods allow to reconstruct the EoS using the model-independent approach. Below, we explain in detail each of the methods.

\subsection{The Distance Modulus of GRBs}
In this case, we are going to reconstruct the EoS using the distance module measures derived from the GRBs.
Much research has been conducted to establish GRBs as standard candles, since there is not only one method to standardize GRBs, for example, the distance calibrations GRBs usually make use of some of the empirical luminosity correlations, such as, $\tau_{lag}-L$, $V-L$ or $E_{p}-E_{iso}$ relations, among others \footnote{The $\tau_{lag}-L$ is a correlation between spectrum lag and isotropic peak luminosity. The $V-L$ is a correlation between time variability and isotropic peak luminosity and the $E_{p}-E_{iso}$ represents a tight correlation between the peak energy of spectrum and isotropic equivalent energy.}.
Nevertheless, the different publications show that the constraints, on the cosmological parameters, using GRBs data are compatible with the results of supernovae Ia, which are currently the best standard candles. For a review of the current status of the standardization methods and details of the empirical relations, see references \cite{basilakos}.

Recently, Demianski et al. \cite{demianski} constructed a sample of GRBs using the 
luminosity distance of supernovae Ia to calibrate the correlation between the peak photon 
energy and the isotropic equivalent radiated energy of the GRBs (this is $E_{p}-E_{iso}$ relation) and, consequently, 
construct a Hubble diagram of GRBs. We use this calibrated sample for our analysis.
The GRBs sample that we use consists of 162 measurements and covers a redshift 
interval between $ 0.03 \leq z \leq 9.3 $ \cite{demianski}.

The theoretical distance module for the GRBs can be defined as follows:

\begin{eqnarray}
 \mu_{th}(z,\theta) = 25 +5log D_{L}(z,\theta),
\end{eqnarray}
where $\theta$ represents the parameters of a given cosmological model and $D_{L}$ 
represents the dimensionless luminosity distance. However, for our calculations, it is more convenient to rewrite this equation to determine the dimensionless comoving distance as follows:
\begin{eqnarray}
 D_{c} = \frac{10^{\frac{\mu_{th}-25}{5}}}{1+z}.
\end{eqnarray}
In the above equation, we have used the fact that when the tricurvature is flat, the comoving distance is related to the luminosity distance by the 
relation\footnote{From here on we use comoving distance 
instead of dimensionless comoving distance for $D_{c}$.}: $D_{L} = (1 + z)D_{c}$, and the comoving distance as a function of the Hubble parameter is defined by the following expression:
\begin{eqnarray}
 D_{c} = \frac{c}{H_{0}}\int^{z}_{0}{\frac{dx}{h(x,\theta)}},
\end{eqnarray}
where $h(z,\theta)$ is the dimensionless Hubble parameter,$\frac{H(z)}{H_{0}}$. In our case, it is given explicitly by:
\begin{eqnarray}
 h^{2}(z,\Omega_{m0}, \Omega_{k}) =  \left\{\Omega_{m0} (1+z)^3+\Omega_{k}(1+z)^2
+(1-\Omega_{m0}-\Omega_k)\exp{\left[3\int_0^z \frac{1+w(z')}{1+z'}\mathrm{d}z'\right]}\right\},
\end{eqnarray}
where $\Omega_{m0}$ and $\Omega_{k} $ represent the matter density parameter and curvature respectively.
In this paper, we assume that $\Omega_{k}=0$, which matches the results of the Planck satellite \cite{aghanim}.
To derive from the previous equation, an expression for EoS, is useful the relation established by Huterer and Turner, among others \cite {Starobinsky, huterer, nakamura}, 
\begin{eqnarray}
 H(z) = \frac{1}{D'_{c}} = \left[(\frac{D_{L}}{1+z})'\right]^{-1},
\end{eqnarray}
being that the “prime” represents the derivative with respect to redshift.
Therefore, we can use this equation together with the equation of $h(z)$ to derive the equation of state as a function of $ D_{c}$ \cite{zunckel}:
\begin{eqnarray}
 w(z)= \frac{2(1+z)(1+\Omega_{k} D_{c}^2)D_{c}''-[(1+z)^2\Omega_{k} D_{c}'^2
+2(1+z)\Omega_{k} D_{c}D_{c}'-3(1+\Omega_{k} D_{c}^2)]D_{c}'}{3{(1+z)^2[\Omega_k+(1+z)\Omega_m]D_{c}'^2-(1+\Omega_{k} D_{c}^2)}D_{c}'}
\end{eqnarray}

Observing the previous equation, we can see that if we have observational data of the comoving distance 
and a method to reconstruct the derivatives of these data, then we can reconstruct the EoS directly 
from the observational data, except that we have to assume values for $\Omega_{m0}$.
This is a characteristic that always occurs when the EOS is reconstructed.

\subsection {The Sandage-Loeb Effect: Simulation of the Hubble Parameter}
In addition to the data of the GRBs, we use simulated data through the Sandage-Loeb effect (SL). 
The SL effect is a purely geometric measurement of the expansion rate of the universe. 
It can detect a change in the redshift of the source in the 
spectra of the Lyman-$\alpha$ forest of distant quasars in the range of $2{\leq}z{\leq}5$, also know as the "redshift desert".
Initially, this effect was proposed by Sandage \cite{sandage} in 1962. However, using the 
technology of the time, a temporary interval of the order of $10^7$ years was required to obtain an appreciable change in the redshift. 
Nevertheless, a reanalysis of the effect made by Loeb \cite{loeb} relaunched the interest in this effect. Loeb argues that 
by using measurements from the Lyman-$\alpha$ forest of high-redshift quasars and with high-spectroscopic 
resolution associated with a 10-m class telescope, it would be possible to directly detect cosmic acceleration 
using time measurements of the order of $10^{1}$ years.
Currently, the CODEX (COsmic Dynamics and EXo-earth experiment)\cite{codex}, which is an optical high spectral resolution 
instrument and proposed for the European Extremely Large Telescope (E-ELT), can possibly detect this effect.

On the other hand, the effect of the atmosphere on experiments such as CODEX is strong and this is the 
reason why the measurements are planned to be obtained in the "redshift desert" 
\footnote{Another possibility is to observe the sign of neutral hydrogen (HI) of galaxies in two different epochs. This experiment has been proposed for the SKA (Square Kilometre Array). 
In particular, the effect of the atmosphere is negligible on measures of the sign of the neutral hydrogen. This allows the SKA to make measurements for $ z <1.65 $ \cite{ska}.}. 
As we are interested in reconstructing the equation of state at high redshift, we are 
going to use the CODEX prescriptions to simulate our data.

We can calculate the expected variation of the redshift of a chosen extragalactic source with time. If we assume  that the source does not have any peculiar velocity, and so any peculiar acceleration, then it has a fixed comoving coordinate. 
In a homogenous and isotropic universe with a FLRW metric, consider that a source emits an electromagnetic wave during an interval 
($t_{s},t_{s}+\delta t_{s}$) and this wave is received by an observer during an interval ($t_{0},t_{0}+\delta t_{0}$), then we can write \cite{weinberg}:

\begin{eqnarray}
\int_{t_{s}}^{t_{0}}{\frac{dt}{a(t)}} = \int_{t_{s}+\Delta t_{s}}^{t_{0}+\Delta t_{0}}{\frac{dt}{a(t)}}.
\end{eqnarray}

If we consider that $\Delta t_{s}$, $\Delta t_{0}<<t_{s},t_{0}$, for example, for a typical light signal have an interval $10^{-14}s$, then the above expression leads to
\begin{eqnarray}
 \frac{\Delta t_{0}}{a(t_{0})} \approx \frac{\Delta t_{s}}{a(t_{s})}.
\end{eqnarray}

Considering that the redshift of the observed source is  defined as $z(t_{0})=\frac{\lambda_{0}-\lambda_{e}}{\lambda_{e}}$, 
we can arrive at the well-known expression
$z(t_{0})+1=\frac{a(t_{0})}{a(t_{e})}$. Thus, the source redshift changes for this interval of time can be write as
\begin{eqnarray}
\Delta z_{e} =z_{e}(t_{0}+\Delta t_{0}) - z_{e}(t_{0})  {\equiv} \frac{a(t_{0}+ \Delta t_{0})}{a(t_{s}+ \Delta t_{s})}- \frac{a(t_0)}{a(t_s)}.
\end{eqnarray}
Expanding the first ratio on the right side of the equation above and considering only the first expansion order, we obtain:
\begin{eqnarray}
\Delta z_{s} = \left[\frac{\dot{a}(t_0)-\dot{a}(t_s)}{a(t_s)}\right]\Delta t_{0},
\end{eqnarray}
where the dot represents the time derivative. Thus, this equation can be rewritten as
\begin{eqnarray}
\Delta z_{s} = \left[H_{0}(1+z_{e}(t_{0})) - H(t_{e})\right]\Delta t_0.
\end{eqnarray}

It is more convenient to express the redshift variation as a spectroscopic velocity shift as
\begin{eqnarray}
\Delta v \equiv \frac{c \Delta z_{s}}{1+z_{s}} = - H_{0}\Delta t_{0} c[1-\frac{h(z)}{1+z}],
\end{eqnarray}
where as previously, $h(z) = \frac{H(z)}{H_{0}}$. 
Therefore, if we can measure $\Delta z_{s} $ or equivalently $\Delta v $,
then we can obtain a measure estimate of the Hubble parameter.

For a simulation of the Hubble parameter data using the SL effect, we used the $\Lambda CDM$ model as the fiducial model,
\begin{eqnarray}
h(z) = \sqrt{\Omega_{m0}(1+z)^3+ \Omega_{r0}(1+z)^{4}+(1-\Omega_{m0}-\Omega_{r0})},
\end{eqnarray}
where we have included a radiation component $\Omega_{r0}$ and assume the values $H_{0}=70 $ km/s/Mpc and $\Delta t_{0}=10^{1}$ years.

The next step was to calculate the $\sigma_{\Delta v}$. 
To estimate the error, we use the prescription indicated by the collaboration CODEX \cite{codex}, 
which establishes that the accuracy with which you can determine $\Delta v$ from the Lyman-$\alpha$ forest can be written as
\begin{equation}
\sigma_{\Delta_{v}} = 1.35 (\frac{S/N}{2370})^{-1}(\frac{N_{QSO}}{30})^{1/2}(\frac{1+z_{QSO}}{5})^{f} \frac{cm}{s},
\end{equation}
where $S/N$ is the signal-to-noise ratio defined per $0.0125 \dot{A}$ pixel. 
$N_{QSO}$ is the number of observed quasars, $z_{QSO}$ represents their redshifts, $f=-1.7$ in the interval $ 2 < z < 4$ 
and  $f=-0.9$ for $z > 4$. The chosen number of quasars were $30$ and the $S/N = 3000$.

To determine the Hubble parameter, we must invert the equation to the spectroscopic velocity shift
\begin{equation}
 H(z) = [H_{0}-\frac{\Delta v}{c \Delta t_{0}} ] (1+z).
\end{equation}

Since error propagation allows us to determine the uncertainty of $H(z)$ as
\begin{equation}
 \sigma_{H} = \frac{1+z}{\Delta t_{0}} \sigma_{\Delta v}
\end{equation}

Therefore, as our simulated data are the Hubble parameter data, we need an expression that allows us to reconstruct the EoS, $w(z)$, 
directly from this observable. For such a case, it is possible to derive one expression directly from equation (4) as \cite{huterer2,seikel}

\begin{eqnarray}
 w(z) = \frac{2(1+z)hh'-3h^{2}+\Omega_{k}(1+z)^{2}}{3[h^{2}-\Omega_{m0}(1+z)^{3}-\Omega_{k}(1+z)^{2}]}.
\end{eqnarray}
Again, it is interesting to note that this equation, 
as well as the previous equation for $w$, requires an external knowledge of the value of the $\Omega_{m0}$ parameter 
and in this case, by equation (16), also of the parameter $H_{0}$.

\subsection{Gaussian Processes}
To perform the reconstruction of the equation of state, we use the non-parametric method of Gaussian processes, 
which is particularly important because it does not assume a specific model of the EoS.
A Gaussian process can be written as:
\begin{eqnarray}
f(x) \sim GP\left(\mu(x),k(x,\tilde{x})\right),
\end{eqnarray}
where the value of $f$ when evaluated at a point $x$ is a Gaussian random variable with 
mean $\mu(x)$. Additionally, the value of the function $f$ is not independent of the value of the 
function $f$ at some other point nearby $\tilde{x}$, but is related 
by the covariance function $k(x,\tilde{x})$ . For our calculations, we assume the exponential
function as a covariance function which is given by
\begin{eqnarray}
k(x,\tilde{x})=\sigma _{f}^{2}\exp \left(-\frac{(x-\tilde{x})^{2}}{2\ell^{2}}\right),
\end{eqnarray}
where $\sigma_{f}$ and $\ell$ are called hyperparameters which are determined by using the likelihood method for the data.
Additionally, this method 
allows to reconstruct the derivative of the data. To implement this method, we use 
the public package GAPP \cite{gapp}. 
For details on the statistical method, reference \cite{rasmussen} may be useful and for applications in cosmology, consider 
reference \cite{seikel}.

\subsection{Methodology}
In an illustrative way, our methodology is summarized in Fig. 1. Initially, we begin with observational data or simulated data. 
These data are analyzed using a non-parametric statistical method. As a result of this statistical analysis, we obtain, 
as in our case, the comoving distance $D_{c}$ and its derivatives $ D'_{c}$ and $ D''_{c}$ 
and the Hubble parameter $h$ and its derivative $h'$. Then, we use a model-independent approach to reconstruct the EoS. To estimate the errors of the reconstruction, we use the Monte Carlo method.

This methodology is quite general because we can assume other non-parametric methods in addition to the Gaussian process method, such as the 
principle component analysis \cite{nesseris,crittenden,clarkson,zhao} or, as in reference \cite{escamilla}, a non-parametric method that 
consists of a combination of the Loess and Simex methods. Analogously, another model-independent approach can be assumed. 
For instance, we can assume a consistency test of the $\Lambda CDM$ \cite{zunckel} or we can use the cosmographic parameters \cite{weinberg}. 
To estimate reconstruction errors, an alternative to the Monte Carlo method is the method of resampling, such as the jackknife method 
\cite{shisa}, among others.

\begin{figure*}[htbp] 
 	\centering
 		\includegraphics[scale=0.370]{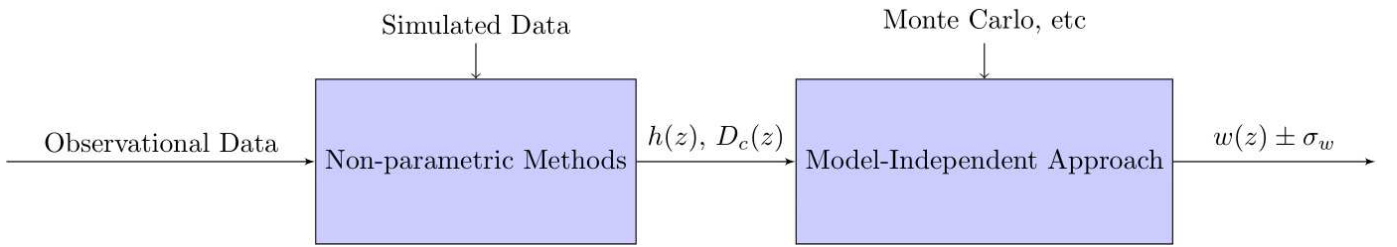}
 	\caption{Schematic summary of the methodology used.}
 	\label{fig:epsilon_01_03}
 \end{figure*}

  \begin{figure*}[htbp] 
 	\centering
 		\includegraphics[scale=0.775]{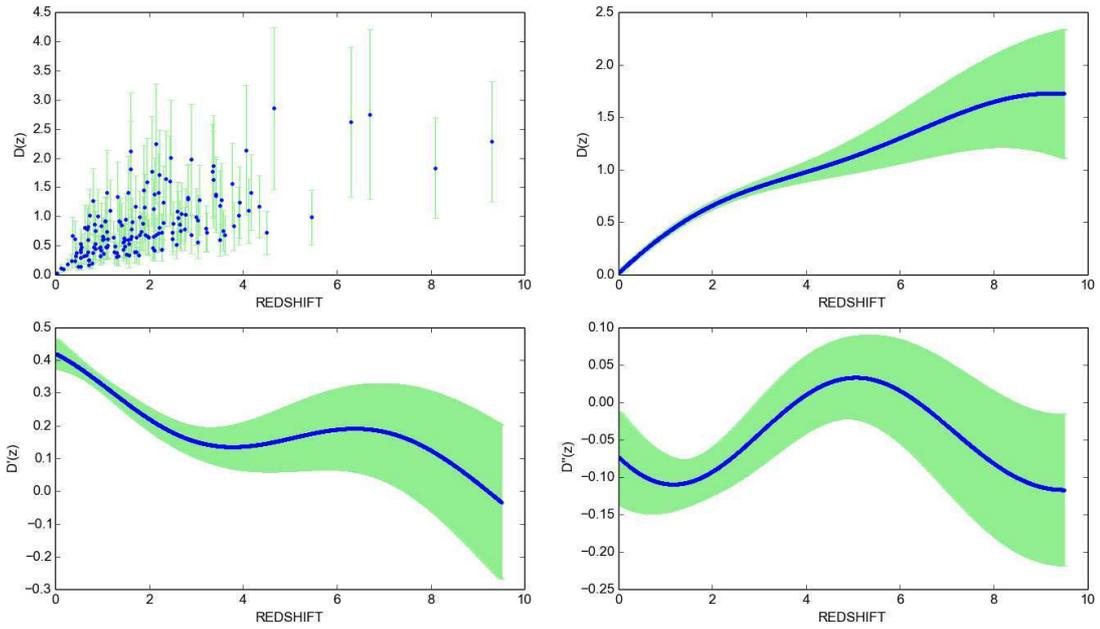}
 	\caption{On the left-hand plot, we shown the 162 Data of the GRBs for the comoving distance. 
 	In the other plots, we show the reconstruction of this observable and its derivatives.}
 	\label{fig:epsilon_01_03}
 \end{figure*}

 \begin{figure*}[htbp] 
 	\centering
 		\includegraphics[scale=0.150]{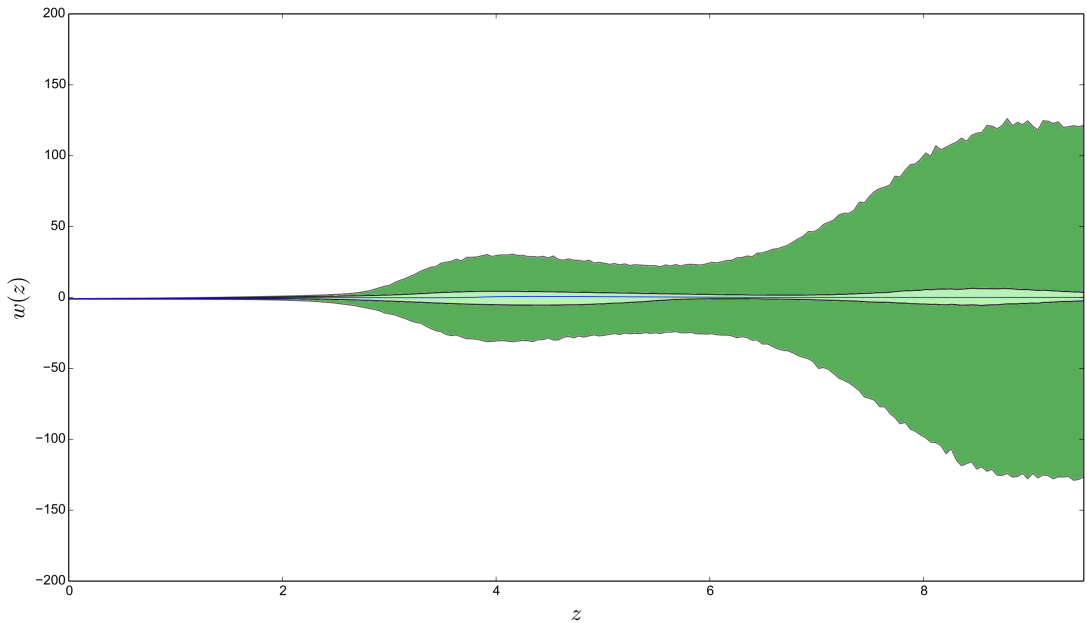}
 	\caption{The cosmic equation of state using the reconstructed data of the comoving distance by the model-independent approach is provided by equation (6).
 	The calculation of uncertainty is determined by Monte Carlo.
 	The reconstruction is done for the intervals: $0.03\leq z \leq 9.3$. We can see that for redshift greater than $z>3$, 
 	the errors increase noticeably. The dependence on the value of the parameter $\Omega_{m0}$ for high redshift are covered by the large errors.}
 	\label{fig:epsilon_01_03}
 \end{figure*}

 \begin{figure*}[htbp] 
 	\centering
 		\includegraphics[scale=0.100]{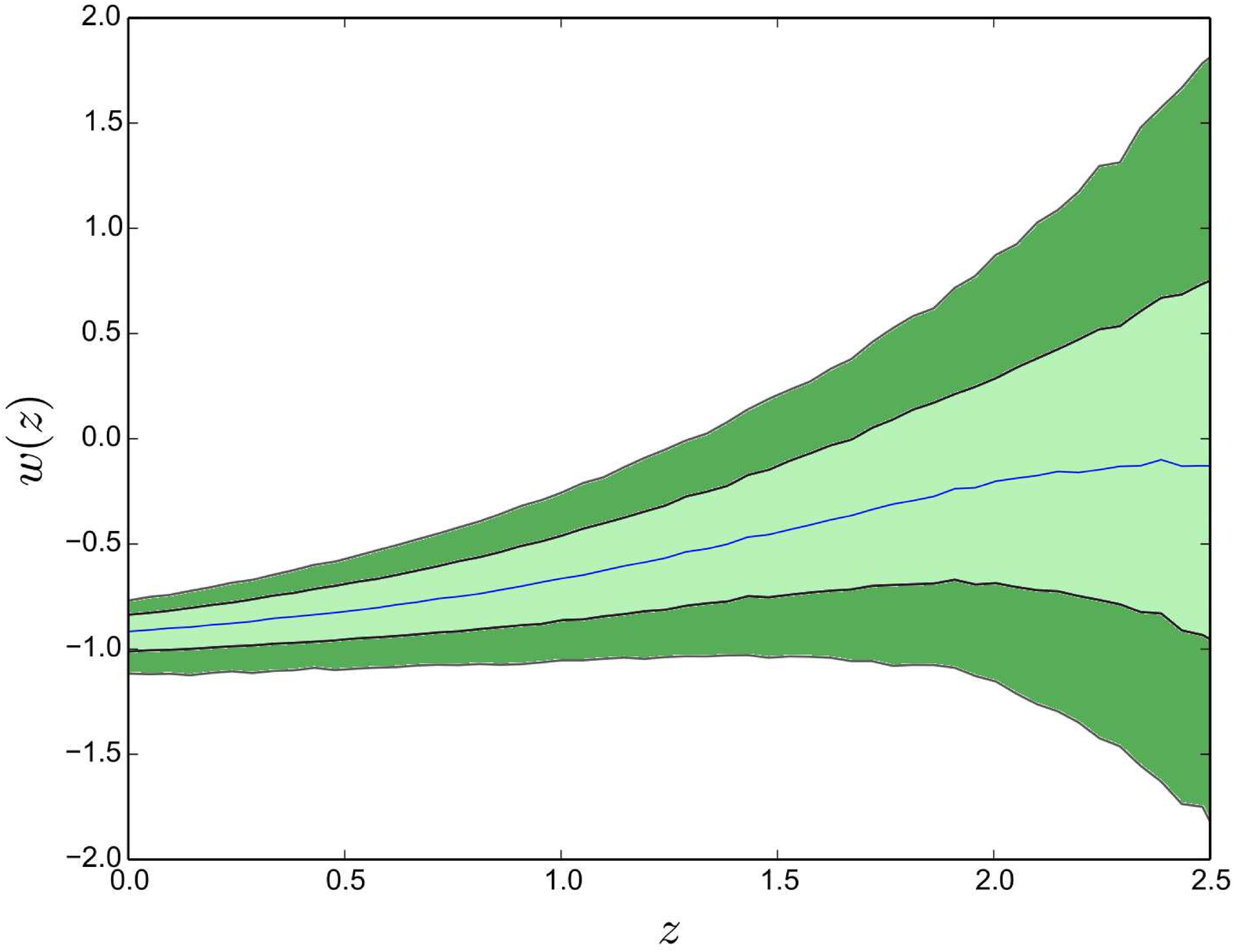}
 		\includegraphics[scale=0.100]{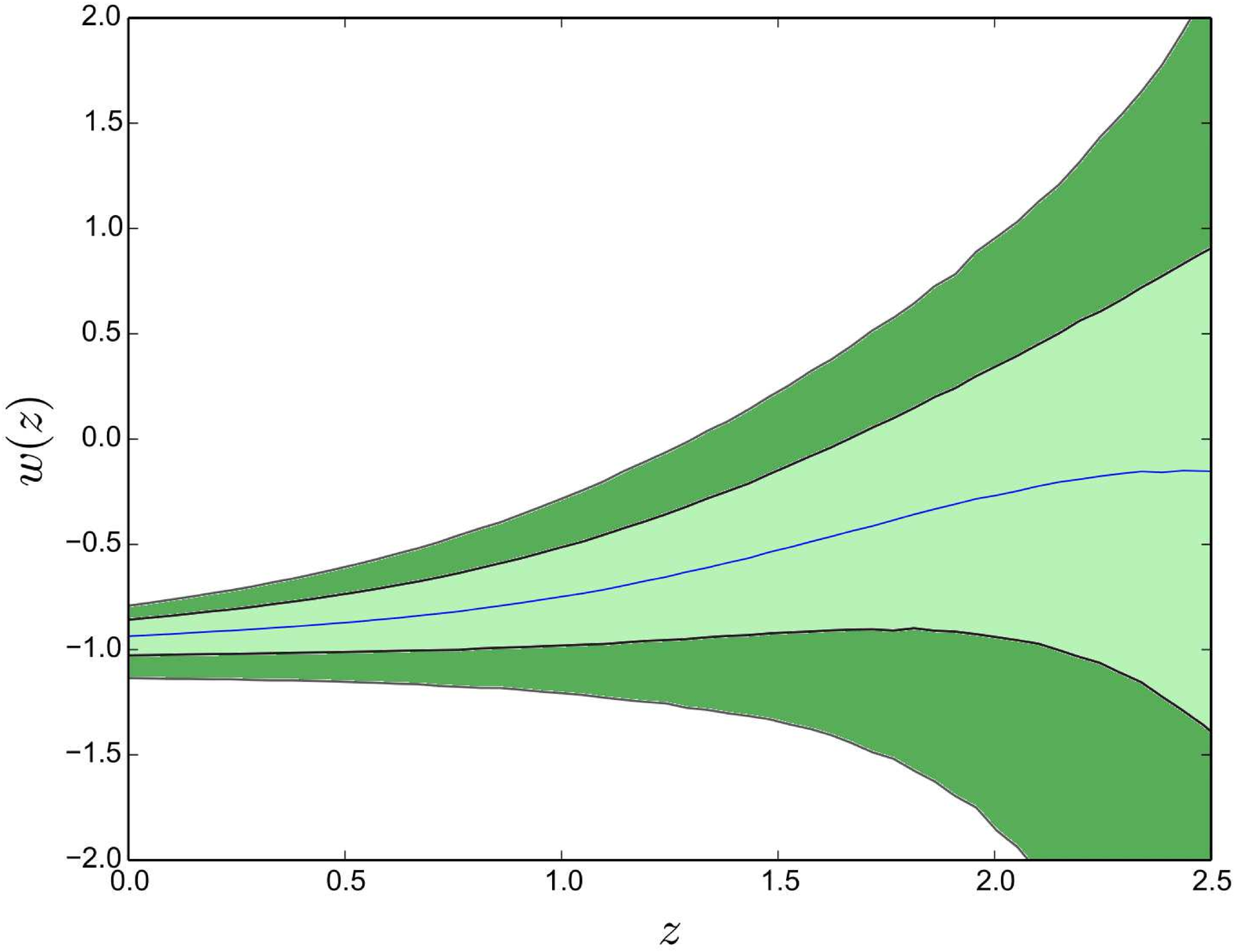}
 		\includegraphics[scale=0.100]{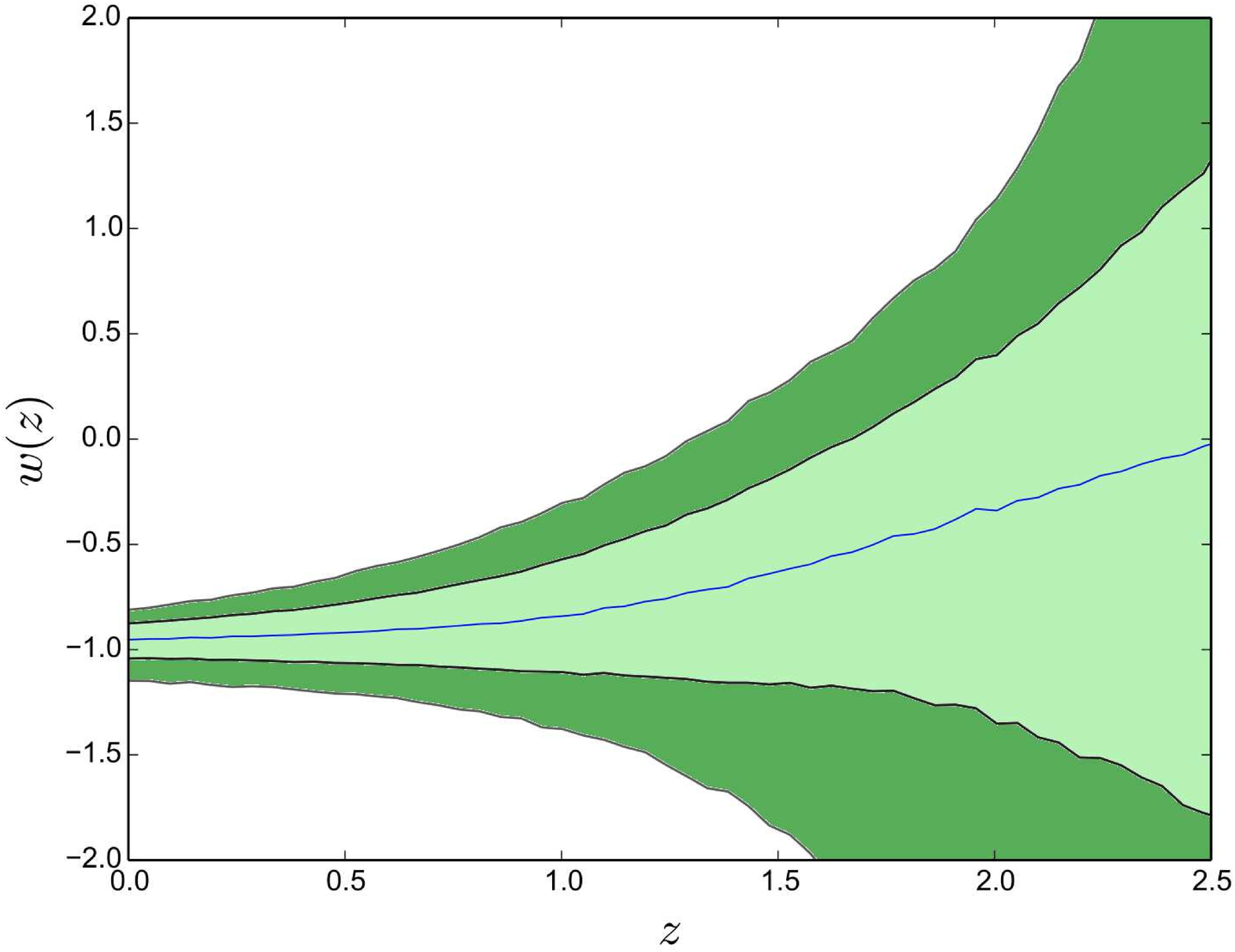}
 	\caption{We show the dependency of the reconstruction of the state equation with the value of $\Omega_{m0}$. 
 	This dependence is evident for minor redshifts that are depicted in the $3$ figures. In the left, $\Omega_{m0}=0.200$; in the centre, 
	$\Omega_{m0}=0.275$ and; to the right, $\Omega_{m0}=0.400$.}
 	\label{fig:epsilon_01_03}
 \end{figure*}
 
 \begin{figure*}[htbp] 
 	\centering
 		\includegraphics[scale=0.775]{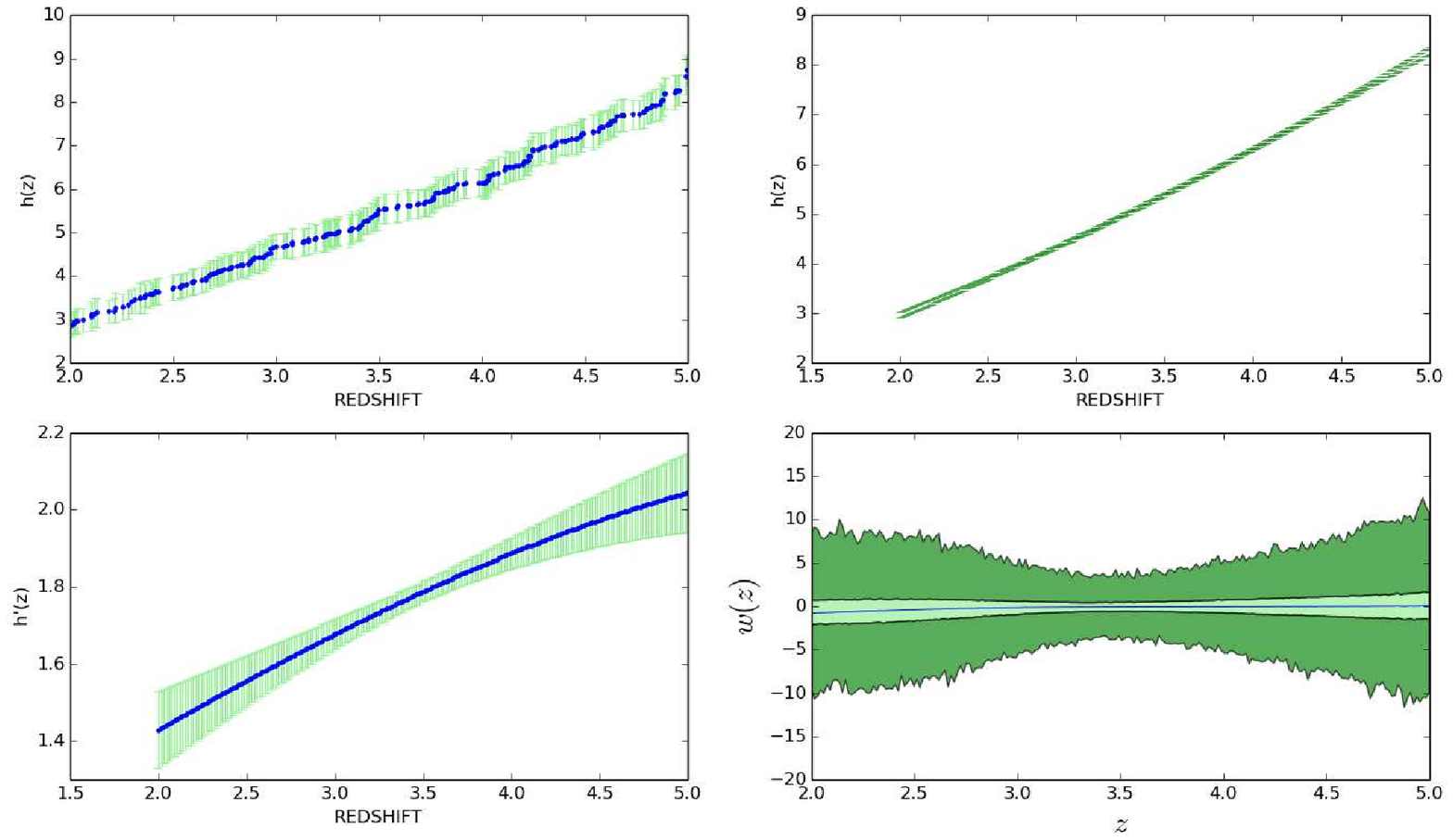}
 	\caption{In the top, we can see the simulated data using the Sandage-Loeb effect and the reconstruction of $h(z)$ using the Gaussian process.
 	In the bottom, we see the reconstruction of the first derivative $h'(z)$ and the forecasted constraints for the cosmic equation of state with $\Omega_{m0}=0.275$.}
 	\label{fig:epsilon_01_03}
 \end{figure*}

\section{Results}
In Fig. 2, we see the results of the reconstruction of the comoving distance and its derivatives 
using the GRBs data, following the methodology presented in section II. In Fig. 3, we see the complete reconstruction of the EoS. 
As observed, the errors increase significantly for high redshifts, especially for values $ z> 2.0 $. 
In Fig. 4, we present the effects of changing the value of $ \Omega_{m0}$.
We see that the effect is very small. What we can emphasize is that in all the cases, for high redshift, 
we observe a tendency of the average value of $w(z)$ for a universe dominated by matter.

In general terms, we can say that the EoS reconstructed with one and two sigmas 
shows a fairly large error propagation and it does not allow us to distinguish between cosmological models. However, 
we have proved that even with large error propagation, it is possible to reconstruct the EoS. This same methodology could be used when 
the number and quality of the GRBs data increases considerably.

In Fig. 5, we can see the result of applying the methodology of the section II.B. We presented
the simulated data for the function $h(z)$  and its derivative $h'(z)$ in a redshift between $2-5$
using the Gaussian processes.
The simulated errors are relatively small when compared to the result of the GRBs. 
For example, compare the vertical scale between Fig. 3. and the reconstruction shown in the lower right section of Fig. 5.
Although they are not real data, we hope that future data will follow this trend and that robust constraints on the observables can be determined. 

On the other hand, given that the SL effect depends on the assumed value of $H_0$, 
we considered the dependence of changing this value in the reconstruction of $w(z)$.
Since the errors are still large, the effect of $H_{0}$ on the EoS is quite negligible.
We used the values of $ H_ {0} = 73.24 \pm 1.74$ km/s/Mpc as measured for Supernovae Ia \cite{citesupernova} and the value $H_{0}=67.51 \pm 0.64$ km/s/Mpc 
as extracted from Planck 2015 TT, TE, EE + lowP + lensing data \cite{aghanim}.

\section{Conclusions}
We performed a reconstruction of the EoS using Gaussian processes with GRBs data as well as a model-independent approach. 
The analysis was carried out by considering different values ​​of the $\Omega_{m0}$, which covers the current observation limits. 
Moreover, we also use the SL effect to simulate the Hubble parameter data in the so-called "redshift desert" using
the flat $\Lambda CDM$ model as a fiducial method with values: $\Omega_{m0} = 0.275$ and $ H_{0} = 70$ km/s/Mpc. The reconstruction  
using the simulated data is significantly better (lower propagation of uncertainty) than those determined by the GRBs data. 
However, these measures are strongly dependent on the underlying model.

The fact that our reconstruction is independent of the assumed cosmological model allows us to observe that in general terms,
the form of the reconstruction of $ w(z)$ must be associated with a continuous function and good behavior, without sudden jumps.
Further, it can be seen that the EoS for high redshift is compatible with a phase dominated by matter (as expected by theory). However, with the current data, it is not possible to specify exactly where this happens, i.e., where $w(z)=0$, due to the uncertainty.

In general, to perform a reconstruction of cosmological observables for high redshift, it is necessary to improve the conjunction 
between statistical methods and cosmology. In this study, we have presented specific cases. However, other 
combinations of statistical methods along with independent approximations can be studied. A good calibration of this 
methodology will allow us to use future data to reconstruct the EoS for high redshifts 
with robustness and high precision, which will help discriminate between competitive cosmological models.

\section*{Acknowledgments}
J.C.F acknowledges financial support from FAPES and CNPq. M. M. M acknowledges the support from FAPES.
A. M. V. T would like to dedicate this research to celebrate the scientific career of Ioav Waga, my PhD advisor who has recently retired from UFRJ.

\label{lastpage}
\end{document}